\documentclass{article}
\usepackage[preprint, nonatbib]{neurips_2020}
\usepackage[utf8]{inputenc}
\usepackage{graphicx}
\usepackage{amsmath}
\usepackage{hyperref}
\usepackage{url}            
\usepackage{booktabs}       
\usepackage{amsfonts}       
\usepackage{nicefrac}       
\usepackage{microtype}
\usepackage{tabularx}
\usepackage{wrapfig}
\usepackage[font=small]{caption}

\graphicspath{ {./figures/} }

\title{Self-supervised edge features for improved Graph Neural Network training}
\author{Arijit Sehanobish\thanks{Equal contribution} \\
  Internal Medicine (Cardiology) and Computer Science\\
 Yale University\\
  \texttt{arijit.sehanobish@yale.edu}\\
\and
Neal G. Ravindra$^*$ \\
Internal Medicine (Cardiology) and Computer Science\\
 Yale University\\
  \texttt{neal.ravindra@yale.edu}\\
  \and 
  David van Dijk\\
  Internal Medicine (Cardiology) and Computer Science\\
 Yale University\\
  \texttt{david.vandijk@yale.edu}
  }

\begin{document}

\maketitle

\begin{abstract}

Graph Neural Networks (GNN) have been extensively used to extract meaningful representations from graph structured data and to perform predictive tasks such as node classification and link prediction. In recent years, there has been a lot of work incorporating edge features along with node features for prediction tasks. One of the main difficulties in using edge features is that they are often handcrafted, hard to get, specific to a particular domain, and may contain redundant information. In this work, we present a framework for creating new edge features, applicable to any domain, via a combination of self-supervised and unsupervised learning. In addition to this, we use Forman-Ricci curvature as an additional edge feature to encapsulate the local geometry of the graph. We then encode our edge features via a Set Transformer and combine them with node features extracted from popular GNN architectures for node classification in an end-to-end training scheme. We validate our work on three biological datasets comprising of single-cell RNA sequencing data of neurological disease, \textit{in vitro} SARS-CoV-2 infection, and human COVID-19 patients. We demonstrate that our method achieves better performance on node classification tasks over baseline Graph Attention Network (GAT) and Graph Convolutional Network (GCN) models. Furthermore, given the attention mechanism on edge and node features, we are able to interpret the cell types and genes that determine the course and severity of COVID-19, contributing to a growing list of potential disease biomarkers and therapeutic targets.

\end{abstract}

\section{Introduction}

Graph neural networks (GNN) have been widely used and developed for predictive tasks such as node classification and link prediction~\cite{wu_comprehensive_2020} and have been shown to learn from any sparse and discrete relational structure in data~\cite{franceschi_learning_2019}. In particular, the use of similarity metrics to construct graphs from feature matrices expands the scope of GNN applications to domains where graph structured data is not readily available~\cite{tenenbaum_global_2000}. 
GNNs typically use message passing, or recursive neighborhood aggregation, to construct a new feature vector for a particular node by collecting its neighbor's feature vectors~\cite{xu_representation_2018,GCN}. However, most GNN schemes do not use edge features in learning new representations of graphical data. 

Recently, edge features have been incorporated into GNNs to harness information describing different aspects of the relationships between nodes~\cite{gong2018exploiting, gao2018edge2vec}. However, there are very few frameworks for creating \emph{de novo} edge feature vectors in a domain agnostic manner. 

In this article, using Graph Attention Networks, we propose a self-supervised learning framework that is applicable to any graphical data, in which the learned edge attention coefficients become a set of edge features. We show that this novel approach  improves GNN performance in downstream node classification tasks and improves training. Our framework is broad in the sense that any available metadata associated with a particular node can be fed into a self-supervised learning framework in this manner to extract edge features.

Given the devastating impact of the coronavirus disease of 2019 (COVID-19) caused by infection of SARS-CoV-2 and the gap in our understanding of the molecular mechanisms of the disease, we sought to focus our efforts on COVID-19 datasets that can generate hypotheses related to these gaps~\cite{yan_interpretable_covidmort,zhong_immunology_2020}. 
Our focus on single-cell transcriptomic data relating to COVID-19 was motivated by recent work showing that Graph Attention Networks are effective at predicting disease states on an individual cell-to-cell basis~\cite{ms_paper}. 
Single-cell RNA sequencing (scRNA-seq) is a technology that yields large datasets comprising many thousands of cells' gene expression in a variety of conditions~\cite{zheng_scrnaseq_2017, scrnaseq_medicine, review_scrnaseq}. However, identifying factors important for determining an individual cell's pathophysiological trajectory or response to viral insult remains a challenge as single-cell data is noisy, sparse, and multi-dimensional \cite{compchallenge, rnacluster}. We reasoned that our framework's improved performance could extract useful insights into the genes and cell types that might be important determinants of COVID-19 severity and SARS-CoV-2 infection.

Our paper makes the following contributions:

\begin{itemize}
    \item \textbf{Creation of edge features using semi-supervised learning} We propose a new approach to obtaining edge features that does not require supervision. From a feature matrix, we use unsupervised Louvain clustering in the graph domain to obtain community labels per node~\cite{louvain}. Then, using these cluster labels, we train a GAT model and use the learned edge attention coefficients as self-supervised edge features. We show that this self-supervised learning framework improves performance and training with two popular GNN architectures. 
    \item \textbf{Use of Forman-Ricci curvature as an edge feature} We apply a novel way of describing the local structure of a graph by weighting an edge according to the common edges that pass through two nodes in the connection. This unsupervised learning of the graph structure can be applied to any graph and may enable archetypal or prototypical analysis of nodes that have high curvature. 
    \item \textbf{Use of a Set Transformer to ingest edge features and enhance interpretability} In our self-supervised framework, we use a Set Transformer to combine edge features with message passing GNN layers in an end-to-end node classification task. 
    \item \textbf{Identification of potential genes and cell types important to SARS-CoV-2 infection and COVID-19 severity} We use attention mechanisms in the GAT layer and the Set Transformer to propose genes and cells that can be important in determining the temporal dynamics of infection and disease severity in COVID-19 patient samples. These genes and cell types may provide potential therapeutic targets or markers of disease severity.
    
\end{itemize}


\section{Related works}

There is a wealth of research on Graph Neural Networks. A significant amount of work has been focused on graph embedding techniques, representation learning and various predictive analyses using node features. There has been recent interest in using edge features, in addition to node features, to improve the performance of Graph Neural Networks~\cite{gong2018exploiting, chen2019utilizing,Abu_El_Haija_2017}. Many real-world graphs already have edge features, such as common keywords between nodes in citation networks or interaction affinities in protein-protein interaction networks. Even for graphs that do not \textit{prima facie} have weighted edges or directions, one can construct edge features that describe the interaction between two nodes using Jaccard similarity, common neighbor metrics such as Adamic Adar or a variety of similarity metrics after embedding the graph. However in this work we use an unique multi--tasking approach in creating these edge features. We hope that this multi--tasking pre--training regime will further the research in meta--learning in the graph domain. 

Since the primary focus of our work is in biological applications, it becomes necessary to be able to interpret the results of our network to inform further study of biology and medicine. One of the most common and popular ways to interpret machine learning models is via Shapley values~\cite{lundberg2017unified} and it's various generalizations~\cite{Michalak_2013}. However Shapley values require independence of features which is hard to guarantee in general. Thus we follow the approach of~\cite{ms_paper,attn_state_space_schaar} in using attention mechanisms for interpretability. Thus even though set2set~\cite{vinyals2015order} is a popular mechanism to encode sets and has been previously used in the graph domain~\cite{vinyal_message_passing, altaetran2016low}, our view is that it is hard to interpret the hidden state of a LSTM. The transformer model~\cite{Attention, lee2018set}, on the other hand, allows us to interpret the results by looking at their attention heads.

GNNs have been used in biomedical research to predict medications, diagnoses, and outcomes from graphical representations of electronic health records~\cite{GRAM}. GNNs have also been used to predict protein-protein and drug-protein interactions and molecular activity~\cite{nguyen_graphdta_2019, chan_drugdiscoveryAI,Harikumar837807,GAT}. However, fewer works attempt to predict pathophysiological state on an individual cell basis. One recent work uses GAT models to predict the disease state of individual cells derived from clinical samples~\cite{ms_paper}. However, their work ignores edge features, which may contain important information regarding cell type and the source from which a particular cell is derived. 
They do not consider multiple disease states or severity nor do they account for the confounding bias of batch effects, which may allow the network to learn a label for an individual cell based on its origin~\cite{review_scrnaseq}. Here, we use the information contained within the dataset and a graph-structure that balances the batches, thus reducing the bias of cell source while preserving biological variation~\cite{bbknn_support}. As such, we believe interpreting our model will provide information that is more biologically relevant than proposed in the previous works. 
To the best of our knowledge, this is the first attempt to apply a Deep Learning model to gain insight into multiple pathophysiological states, merging time-points, severity, and disease-state prediction into a multi-label node classification task from single-cell data using edge features.

\section{Methods}

Our work consists of three parts: (1) creating new edge features via self-supervised learning and local curvature; (2) using the edge features for downstream node classification tasks in an inductive setting by encoding edge features via a Set Transformer and node features via message passing GNN layers; (3) using attention coefficients to interpret our results and provide insights into our datasets.

\subsection{Creating edge features via self-supervised learning}

We use a combination of semi-supervised and unsupervised learning, and local graph geometry to create new edge features. We then concatenate these vectors to create a new edge feature vector.

\subsubsection{Using semi-supervised learning to create new edge features}
We use semi-supervised learning to create two types of edge features. \\
\textbf{Using labels from unsupervised clustering in the spectral domain:} Our first type of edge features rely on community detection by optimizing modularity. We use the Louvain clustering algorithm to assign nodes to different communities, as it has been widely used in single-cell data analysis, though it can be applied to any graph~\cite{louvain, rnacluster}. We then use a GAT model to predict the community labels per node.
Finally, we extract the edge attention coefficients from the first layer using equation~\ref{eqn:new_edge_feat}. In this way we get an $h$ dimensional vector, where $h$ is the number of heads, which becomes $h$ edge features. \\
\textbf{Using other node metadata:} 
Some graphs may have node labels that are not of interest for a particular task. For example, in single-cell data, cells from different patients, experiments, or sources are referred to as ``batches" and are pooled into one dataset.
In our datasets, we use the batch label for an individual cell as additional input for self-supervised learning. Using the same method as before, we construct an $h$-dimensional vector from the GAT model after batch label prediction. 

\subsubsection{Forman-Ricci curvature}
We now use the internal geometry of the graph to create our next edge feature. We use the discrete version of the Ricci curvature as introduced by Forman~\cite{Forman} and discussed in~\cite{Samal_2018}.
\begin{equation}\label{forman-ricci}
Ric_{F}(e) := \omega(e) \bigg(\frac{\omega(v_1)}{\omega(e)} + \frac{\omega(v_2)}{\omega(e)} - \sum_{e_{v_1} \sim v_1, e_{v_2} \sim v_2}\bigg[\frac{\omega(v_1)}{\sqrt{\omega(e)\omega(e_{v_1})}} + \frac{\omega(v_2)}{\sqrt{\omega(e)\omega(e_{v_2})}}\bigg]\bigg)
\end{equation}
where $e \in \mathbb{E}$ connecting nodes $v_1$ and $v_2$, $\omega(e)$ is the weight of the edge $e$ which we take to the Euclidean distance in the PCA space, $\omega(v_i)$ is the weight of the node which we take to be $1$ for simplicity and $e_{v_i} \sim v_i$ is the set of all edges connected to $v_i$ and \textit{excluding} $e$. This is an intrinsic invariant that captures the local geometry of the graph and relates to the global property of the graph via a Gauss-Bonnet style result~\cite{watanabe2017combinatorial}.

\subsubsection{Create edge features via node2vec}
We use a popular embedding method called node$2$vec~\cite{grover2016node2vec} to embed the nodes in a $d$ dimensional space. We then calculate the dot product between these node embeddings as a measure of similarity. However to be consistent with our other methods we only compute the dot product between the nodes which share an edge. 
node2vec embeddings preserve the local community structure of a graph, which we expect should provide information to enable enhanced discriminability between nodes, as previously suggested~\cite{khosla_comparative_2019}. 


\subsection{Models}
In this subsection we describe our model, which consists of two components: (1) A Set Transformer and (2) Message passing GNNs, like GCN or GAT layers. \\
Thus our model is quite general and is readily applicable with a wide variety of architectures. Since we use a GAT model to do feature extraction to create the edge features, we will describe the GAT model in detail below. 
\subsubsection{Set Transformer}
We use a Set Transformer as in~\cite{lee2018set}. The Set Transformer is permutation invariant so it is an ideal architecture to encode sets. The building block of our Set Transformer is the multi-head attention, as in \cite{Attention}. Given $n$ query vectors $Q$ of dimension $d_q$, a key-value pair matrix $K \in \mathbb{R}^{n_v \times d_q}$ and a value matrix $V \in \mathbb{R}^{n_v \times d_v}$ and, assuming for simplicity that $d_q = d_v=d$, then the attention mechanism is a function given by the following formula: 
\begin{equation}
    \text{att}(Q,K,V) := \text{softmax}(\frac{QK^{T}}{\sqrt{d}})V
\end{equation}
This multihead attention is computed by first projecting $Q,K,V$ onto $h$ different $d^{h}_q, d^{h}_q, d^{h}_v$ dimensional vectors where, for simplicity, we take $d^{h}_q = d^{h}_v = \frac{d}{h}$
\begin{equation}
\text{Multihead}(Q, K, V) := \text{concat}(O_1, \cdots , O_h)W^O
\end{equation}
where 
\begin{equation}
O_j = \text{att}(QW_{j}^{Q}, KW_{j}^{K},VW_{j}^{V}) \end{equation}
and $W_{j}^{Q}, W_{j}^{K},W_{j}^{V}$ are projection operators of dimensions $\mathbb{R}^{d_q \times d^{h}_q}, \mathbb{R}^{d_q \times d^{h}_q}$ and $\mathbb{R}^{d_v \times d^{h}_v}$ respectively and $W^O$ is a linear operator of dimension $d \times d$. Now, given a set $S$, the Set Transformer Block (STB) is given the following formula: 
\begin{equation}
    STB(S):=  \text{LayerNorm}(X + rFF(X))
\end{equation}
where 
\begin{equation}
X = \text{LayerNorm}(S + \text{Multihead}(S, S, S))
\end{equation}
rFF is a row-wise feedforward layer and LayerNorm is layer normalization~\cite{ba2016layer}.
\\
A Set Transformer takes as input a $3$d tensor of the form [batch, seq-length, input-dim] and outputs $3$d tensor of the form [batch, seq-length, output-dim], i.e. it outputs sets of the same size as the input sets. If, for a batch $b_i$, the set transformer outputs a set of the form \{$w_{i1},....w_{ij}$\}, we modify the output of the transformer to a fixed length vector 
\begin{equation}
    w_{i} : = \sum_{j}\lambda_{j}w_{ij}
\end{equation}
where $\lambda_{j}$ are learnable weights. This step is necessary for us because our downstream tasks require vectors of fixed length. 

\subsubsection{Graph Attention Network}
We use the popular Graph Attention Network (GAT) for extracting features from our auxiliary tasks. We follow the exposition in~\cite{GAT}. The input to a GAT layer are the node features, $\mathbf{h} = \{ h_1, h_2, . . . , h_N \}$, where $h_i \in \mathbb{R}^F $, $N$ is the number of nodes, and $F$ is the number of features in each node. The layer produces a new set of node features (of possibly different dimension $F'$) as its output, $\mathbf{h'} = \{h'_1,h'_2,....h'_N \}$ where $h'_i \in \mathbb{R}^{F'}$.  The heart of this layer is multi-head self-attention like in~\cite{Attention, GAT}. Self-attention is computed on the nodes
 \begin{equation}\label{eqn:feed_forward}
     a^{l} : \mathbb{R}^{F'} \times \mathbb{R}^{F'} \rightarrow \mathbb{R}
 \end{equation}
that computes attention coefficients, where $a$ is a feedforward network.
\begin{equation} \label{eqn:unnormalized_attn}
e^{l}_{ij} = a^{l}(\mathbb{W}^{l}h_{i}, \mathbb{W}^{l}h_j )
\end{equation}
where $\mathbb{W}^{l}$ is a linear transformation and also called the weight matrix for the head $l$.
We then normalize these attention coefficients. 
\begin{equation}\label{eqn:att_coeff}
\alpha^{l}_{ij} = \text{softmax}_{j} (e^{l}_{ij} ) = \frac{\text{exp}(e^{l}_{ij} )} {\sum_ {k \in \mathcal{N}_{i}} \text{exp}(e^{l}_{ik})}
\end{equation}
where $\mathcal{N}_i$ is a $1$-neighborhood of the node $i$. The normalized attention coefficients are then used to compute a linear combination of the features corresponding to them, to serve as the final output features for every node (after applying a nonlinearity, $\sigma$):
\begin{equation} \label{eqn:new_feat}
h^{l}_{i} = \sigma \bigg( \sum_{j \in \mathcal{N}_i} \alpha^{l}_{ij}\mathbb{W}^{l}h_{j} \bigg).
\end{equation}
We concatenate the features of these heads to produce a new node feature, $h'_{i} := \left\vert\right\vert h^{l}_{i}$.\\
However, for the final prediction layer, we average the representations over the heads and apply a logistic sigmoid non-linearity. Thus the equation for the final layer is: 
\begin{equation}
    h'_{i} = \sigma \bigg(\frac{1}{K}\sum_{l=1}^{K} \sum_{j \in \mathcal{N}_i} \alpha^{l}_{ij}\mathbb{W}^{l}h_{j} \bigg).
\end{equation}
where $K$ is the number of heads. \\
Our new edge features $\Lambda_{ij}$ for the node $e_{ij}$ are created by concatenating the $\alpha^{l}_{ij}$ across all heads, i.e. 
\begin{equation}\label{eqn:new_edge_feat}
\Lambda_{ij} := \lvert\rvert_{l=1}^{K} \alpha^{l}_{ij}
\end{equation}
Thus we end up with a $K$-dimensional edge feature by this method.

\subsubsection{Our model}

\begin{figure}[h]
  \centering
  \includegraphics[width=0.95\textwidth]{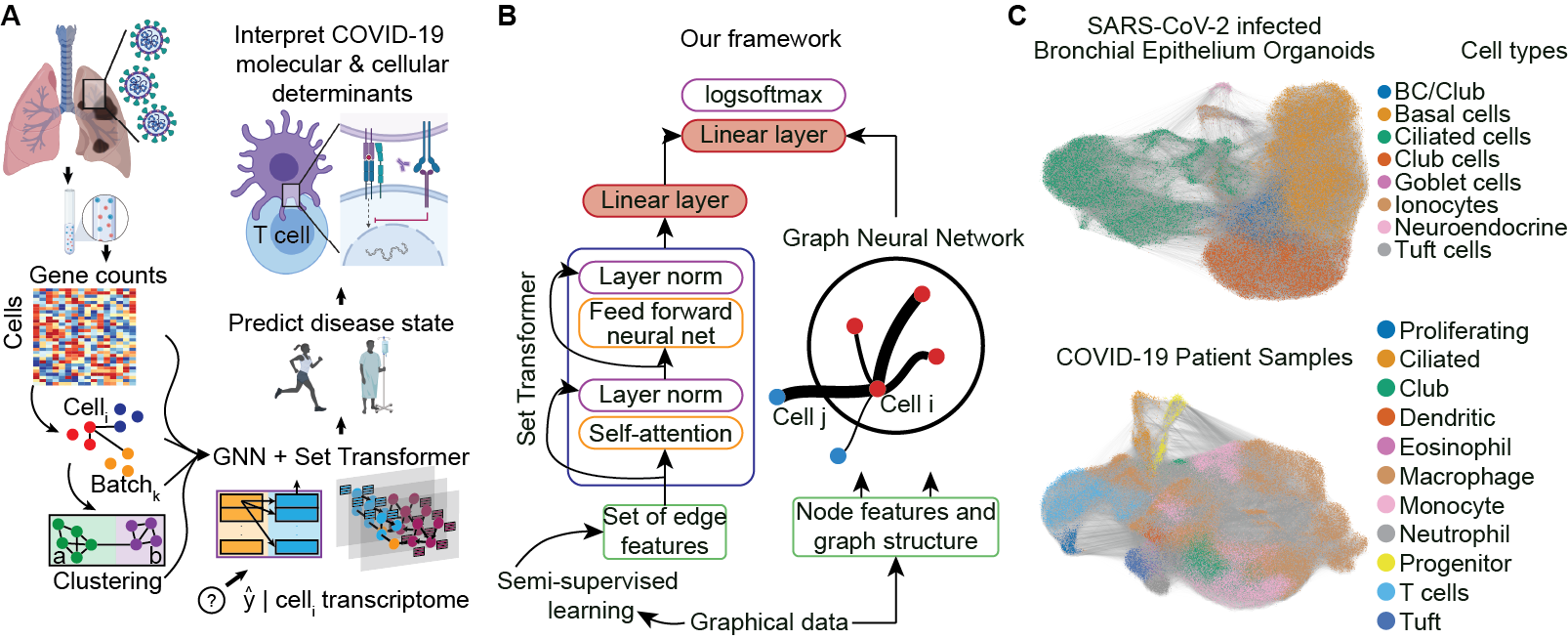}
  \caption[0.95\textwidth]{Our framework and datasets of interest. (\textbf{A}) Overview of our approach with respect to gaining molecular and cellular insights into COVID-19. (\textbf{B}) Our framework and models' architecture, integrating edge features with GNNs via a Set Transformer. (\textbf{C}) Graphical data used, showing cell types for each cell and edges in a node-feature, dimension-reduced embedding.}
  \label{fig:model}
\end{figure}

In this section we will describe our model that combines edge features, obtained as described above, with node features for our main node classification tasks. We use a message passing networks to encode the node features. For example, for all of our experiments, we use either two GCN or two GAT layers. In the case of the GAT layers, we concatenate the representations obtained by different heads. For each node $i$, we construct a set $S_{i}: = \{ e_{ij}: j \in N_{i} \}$, where $e_{ij}$ is the vector representing the edge features of the edge connecting nodes $i$ and $j$.  We then encode this set, $S_{i}$, which we call the edge feature set attached to the node $i$ via our modified Set Transformer. This fixed length vector is concatenated with the node representation obtained after the second GCN or GAT layer. We call this new representation an enhanced node feature vector. This enhanced node feature vector is then passed through a dense layer with a logistic sigmoid non-linearity for classification. Figure \ref{fig:model}A describes our model architecture which uses edge features, connectivity, and node features for node classification in an inductive setting.


\begin{table}[h]
\centering
\caption{Dataset description showing train/val/test splits.}
\label{tab:tab1_datasets}
\resizebox{\textwidth}{!}{%
\begin{tabular}{@{}cccc@{}}
\toprule
Datasets & \begin{tabular}[c]{@{}c@{}}SARS-CoV-2 \\ infected organoids\end{tabular} & COVID-19 patients & Multiple sclerosis patients \\ \midrule
\# Nodes         & 54353/11646/11648     & 63486/13604/13605     & 53280/19980/11988       \\
\# Node features & 24714                 & 25626                 & 22005                   \\
\# Edges         & 1041226/230429/228630 & 2746280/703217/707529 & 6871820/2635746/1602662 \\
\# Edge features & 18                    & 18                    & 18                      \\
\# Classes & 7                    & 3                    & 2                      \\ \bottomrule
\end{tabular}%
}
\end{table}

\section{Experiments}

We validate our model on the following scRNA-seq datasets: 
\begin{itemize}
    \item 4 human bronchial epithelial cell cultures or ``organoids" that were inoculated with SARS-CoV-2 and co-cultured for 1, 2, and 3 days post-infection~\cite{hbec}.
    \item Bronchoalveolar lavage fluid samples from 12 patients enrolled in a study at Shenzen Third People's Hospital in Guangdong Province, China of whom 3 were healthy controls, 3 had a mild or moderate form of COVID-19 and 6 had a severe or critical COVID-19 illness~\cite{liao_single-cell_2020}.
    \item Blood and CSF samples from 13 patients of whom 6 were healthy controls and 7 had multiple sclerosis, a neurological disease~\cite{ms_paper}.
\end{itemize}

\begin{table}[h]
\centering
\caption{Experimental tasks}
\label{tab:tab2_tasks}

\resizebox{\textwidth}{!}{%
\begin{tabular}{l c c c c} 
\toprule
Task & \begin{tabular}[c]{@{}c@{}}SARS-CoV-2\\ infected organoids\end{tabular} & COVID-19 patients & Multiple sclerosis patients &  \\ \midrule
Louvain cluster ID     & Cell type               & Cell type                   & Cell type                &  \\
Batch or node metadata & Culture sample ID       & Patient ID                  & Patient ID + sample type &  \\
Inductive prediction   & Timepoint and infection & No, Mild, or Severe Disease & MS or Healthy                      &  \\ \bottomrule
\end{tabular}
}
\end{table} 

See table~\ref{tab:tab1_datasets} for a summary of our datasets. For all the datasets we create a Batch-Balanced kNN graph to remove the confounding bias of experimental or sequencing differences between samples~\cite{polanski_bbknn_2019}. For more details about data pre-processing and graph construction from single cell data, please refer to the supplementary material. Table~\ref{tab:tab2_tasks} details all the tasks that we perform on our datasets. \\
\\
\textbf{Auxiliary tasks :} We first describe our auxiliary tasks which we devise to create new edge features. We cluster our datasets using Louvain clustering~\cite{louvain}, and annotate these clusters as ``cell types," as commonly done in single-cell analysis~\cite{rnacluster}. Then, we use a $2$-layer GAT with $8$ attention heads in each layer to predict the cell type label. We extract the edge attention coefficients from the first layer of our trained model as edge features. Thus we get an $8$-dimensional edge feature vector by equation~\ref{eqn:new_edge_feat}. All of our biological datasets have a batch ID associated to it, i.e. some metadata that keeps track of the origin of the cell. We use the same method as before to create another $8$-dimensional edge feature vector. More details and results about the auxiliary tasks can be found in the supplementary material. \\
\\
\textbf{Main tasks :} Our final task is node label prediction in an inductive setting, as shown in \ref{tab:tab2_tasks}. All the results shown are from the test set and our model's performance is reported in table~\ref{tab:tab3_main}. Our model outperforms the baseline models by a significant margin. We note that our results on the MS dataset differ from the results as reported in~\cite{ms_paper} since we use a different graph kernel (BB-kNN kernel~\cite{bbknn_support}), which reduces bias due to technical measurement artifacts of the data. We also calculated the p-value (Welch's t-test) between our model and the baseline GAT and GCN models. The p-value was $<.001$ for all the experiments showing that our models are a significant improvement over the baseline.

\begin{table}[h]
\centering
\caption{Results of inductive tasks on single-cell datasets showing accuracy and $95\%$ confidence intervals.}
\label{tab:tab3_main}
\resizebox{\textwidth}{!}{%
\begin{tabular}{@{}ccccc@{}}
\toprule
Models &
  \begin{tabular}[c]{@{}c@{}}SARS-CoV-2\\ infected organoids\end{tabular} &
  \begin{tabular}[c]{@{}c@{}}COVID-19\\ patients\end{tabular} &
  \begin{tabular}[c]{@{}c@{}}Multiple sclerosis \\ patients\end{tabular} &
  \begin{tabular}[c]{@{}c@{}}P \\ (Welch's t-test)\end{tabular} \\ \midrule
GCN                        & 65.43 (65.21-65.65)          & 89.26 (89.06-89.47)          & 73.23 (72.93-73.53) & -      \\
GCN + Edge Features (Ours) & \textbf{81.61 (79.34-83.87)} & \textbf{92.84 (91.95-93.74)} & \textbf{85.06 (83.85-86.28)} & $<0.001$ \\
GAT                        & 73.10 (70.93-75.27)          & 92.25 (91.27-93.24)          & 73.03 (72.22-73.83) & -      \\
GAT + Edge Features (Ours) & \textbf{82.95 (81.75-84.15)} & \textbf{95.12 (94.02-96.22)} & \textbf{85.03 (84.34-85.72)} & $<0.001$ \\ \bottomrule
\end{tabular}%
}
\end{table}

\begin{figure}[h]
  \centering
  \includegraphics{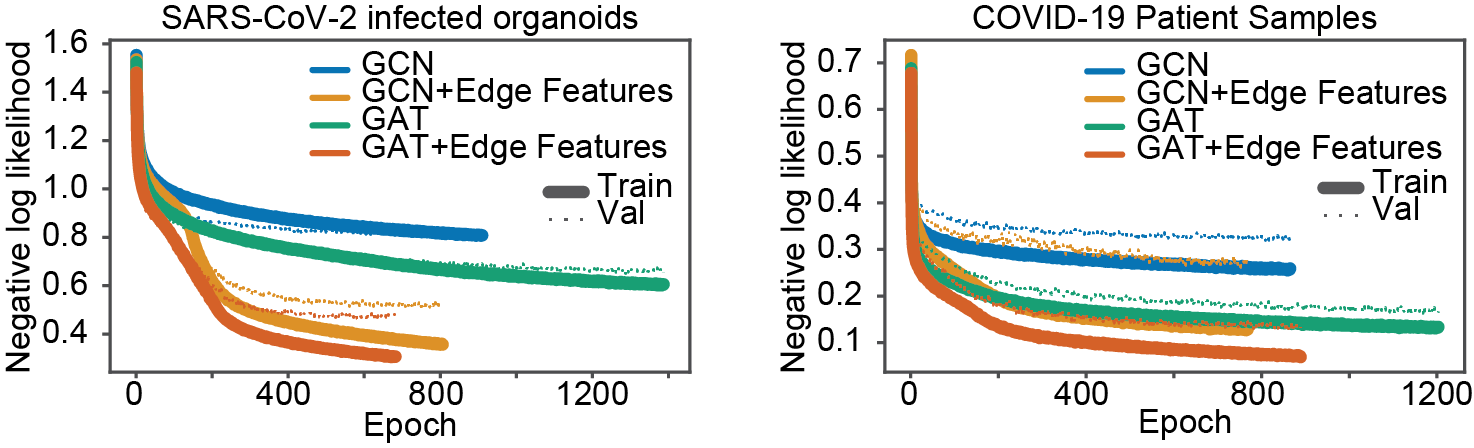}
  \caption[0.9\textwidth]{Negative log likelihood losses per epoch for training models for our COVID-19 datasets.}
  \label{fig:loss}
\end{figure}

Other than improved performance we found that our model trains faster than the baseline GCN and GAT models. We compared the loss per epoch for baseline GCN and GAT models versus our models. Broadly, our model achieves significant and consistently lower loss per epoch and requires fewer epochs to train, as shown in Figure~\ref{fig:loss}. 


\begin{figure}[h]
  \centering
  \includegraphics[width=0.95\textwidth]{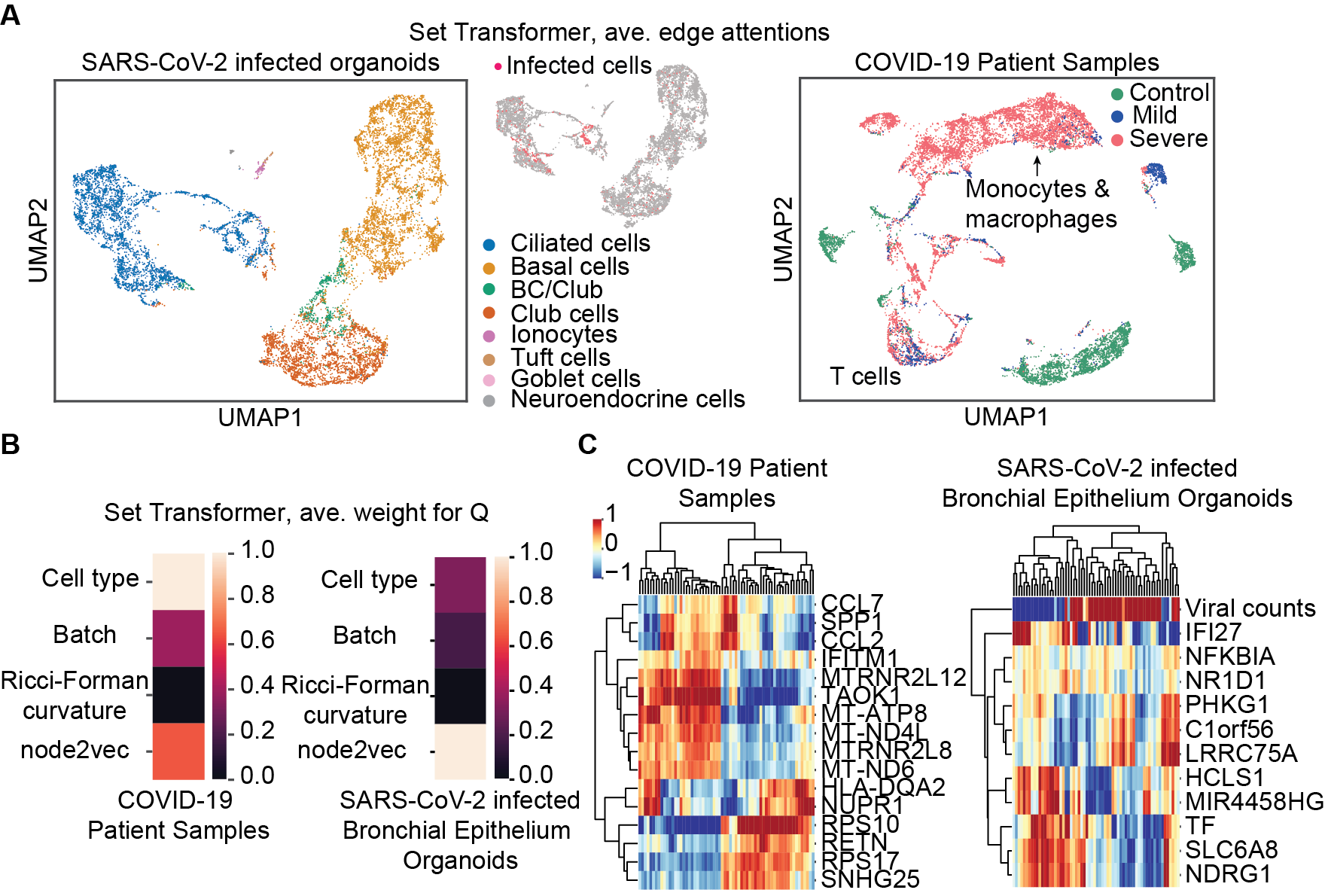}
  \caption[0.95\textwidth]{Model interpretability to generate hypotheses for genes and cells important to COVID-19 severity. (\textbf{A}) Embedding learned from graphs extracted from average edge attentions across Set Transformer output dimensions, showing cell type or condition per cell. (\textbf{B}) Relative importance of crafted edge features in disease state prediction tasks, averaged across the query layer from the Set Transformer. (\textbf{C}) Top 5 important gene features for each GAT head, colored by normalized, learned weights.}
  \label{fig:interpretability}
\end{figure}
\section{Model Interpretability}
In addition, we extract the learned weights from the GAT layer to investigate our model's feature saliency with respect to gene importance in predicting SARS-CoV-2 infection and COVID-19 severity. In predicting COVID-19 severity from patient samples, our model gives high weight to genes involved in the innate immune system response to type I interferon (CCL2, CCL7, IFITM1),  regulation of signaling (NUPR1, TAOK1, MTRNR2L12), a component of the major histocompatibility complex II (HLA-DQA2), which is important for developing immunity to infection, and a marker of eosinophil cells, which are involved in fighting parasites (RETN). In predicting SARS-CoV-2 infection, our model finds saliency in counts of viral transcript, which is encouraging. In addition, to predicting SARS-CoV-2, genes involved in inflammatory response and cell death (NFKBIA) and interferon signaling (IFI27) appear to be important, as does signaling, which may provide clues as to the dynamic response to SARS-CoV-2 infection in the lung's airways (IFI27, HCLS1, NDRG1, NR1D1, TF). 

Using the edge attentions from the Set Transformer, we construct a new graph and perform unsupervised clustering and manifold learning~\cite{umap,louvain}. We obtain distinct cell clusters of SARS-CoV-2 infected cells which are also segregated by cell type. These cells may have unique behaviors that warrant further analysis. The learned embedding for the organoids dataset highlights that our model segregates infected ciliated cells, which is the reported  SARS-CoV-2 cell tropism, validating our models' interpretability~\cite{hbec}. In predicting COVID-19 severity, it is interesting that our model learns to mix macrophages and monocytes in a predominantly severe patient cell cluster while cells derived from mild and severe patients are mixed in a T cell cluster. Monocytes derived from macrophages are thought to be enriched in severe COVID-19 cases and T cells are proposed targets for immune checkpoint therapy of COVID-19, despite lack of understanding as to the transcriptional differences between mild and severe COVID-19 illness~\cite{checkpointinhibitors, liao_single-cell_2020, iwasaki2020}. Lastly, our models find that genes involved in type I interferon signaling are important in predicting both COVID-19 severity and susceptibility to SARS-CoV-2 infection. Interferon signaling is involved in pro-inflammatory immune responses and it is suspected that type I interferon signaling may cause immunopathology during SARS-CoV-2 infection leading to critical illness~\cite{hbec, iwasaki2020}.

\section{Discussion}

We achieved significant performance enhancements using self-supervised edge features when comparing two popular GNN architectures, GCN and GAT models, to our architecture that builds on those models. This suggests that using edge features derived from self-supervised learning and local graph information, with no requirement for hand-crafted edge features, can improve graph neural network performance on challenging node classification tasks. Our models are simple, easy to train and can be used with various graph neural architectures and our edge creation framework is applicable to any graphical data. This flexibility may benefit training and performance, as we show with three biological datasets, but will also expand interpretability to local geometry of the graph using Forman-Ricci curvature. We anticipate that in the future, this metric will help with local explanations of decision boundaries in GNNs. Finally as a future direction we hope to pursue our multi--tasking approach for meta-learning in the graph domain.

Our model allows us to gain insights into the cell tropism of SARS-CoV-2 and to elucidate the genes and cell types that are important for predicting COVID-19 severity. 
We are encouraged to find that genes involved in regulating the immune system are important for predicting SARS-CoV-2 infection and COVID-19 severity. Given the inclusion of edge features and cell types in our model, we are also encouraged that we identified clusters of cells that may be involved in immunopathology~\cite{iwasaki2020, hbec}. Further study into the interaction partners and the subtle transcriptional differences between the cells and cell types we identified may provide complementary hypotheses or avenues for therapeutic intervention to mitigate the impacts of COVID-19.

\section{Broader Impact}

The impact of the COVID-19 pandemic is tragic and its extent is still unknown. 
Here, we attempt to bring accurate disease state prediction to a molecular and cellular scale so that we can identify the cells and genes that are important for determining susceptibility and resistance to SARS-Cov-2 infection and severe COVID-19 illness via interpreting our models.
To the best of our knowledge, no deep learning method can perform as well as we have on predicting multiple disease-states for a single-cell sample. Typically, biologists rely on identifying cells by clustering and dimensionality reduction and compare their differential gene expression to identify molecular determinants of disease. However, this is often done without checking if the differences are meaningful or \emph{predictive}, which we do here. In addition, identifying the cells, cell types, and genes that are important for SARS-CoV-2 infection and COVID-19 severity contributes a long list of potential biomarkers for disease state diagnosis and therapeutic targets for further investigation.
However, there are many caveats to our study. While we achieve good performance with our models, model interpretability in artificial neural networks does not have a strong theoretical basis, and any proposed features should merely be thought of as putative hypotheses into the mechanisms of viral insult. In addition, cells in the COVID-19 patient and MS patient datasets are derived from a relatively small patient population, albeit large for single-cell or clinical studies. While we, for the first time, attempt to limit this bias by using a batch-balanced kNN graph, which we also think makes it more likely that our framework learns from biological variability, we remain vulnerable to the idiosyncrasies of the samples. Thus, any potential feature that identified as important for prediction should only be considered meaningful after extensive experimental validation. We are not medical professionals so we do \emph{NOT} claim that interpretation of our model will bear any fruit. Rather, we hope that the approach of seeking excellent and state-of-the-art predictive results on disease states at single-cell resolution will enhance study of biology and medicine and potentially accelerate our understanding of critical diseases during crises like the COVID-19 pandemic.  

\section*{Acknowledgements}
We acknowledge the Yale Center for Research Computing for our use of their High Performance Computing infrastructure. We thank Mia Madel Alfajaro and Craig B. Wilen for generating the SARS-CoV-2 infected organoids dataset and sharing the data with us. We also thank Jenna L. Pappalardo and David A. Hafler for generating the MS patients dataset and sharing the data with us. 
\nocite{*}
\bibliographystyle{unsrt}
\bibliography{ref}
\newpage
\appendix





\section{Data pre-processing}

\subsection{Feature matrix preparation}

Prior to graph creation, all samples were processed with the standard single-cell RNA-sequencing pre-processing recipe using Scanpy~\cite{Scanpy,satija2015}. Cells and genes for the MS dataset were pre-processed as described in~\cite{ms_paper}. For the SARS-CoV-2 infected organoids and COVID-19 patients datasets, genes expressed in fewer than 3 cells and cells expressing fewer than 200 genes were removed but, to allow for characterization of stress response and cell death, cells expressing a high percentage of mitochondrial genes were not removed. For all single-cell datasets, transcript or "gene" counts per cell were normalized by library size and square-root transformed. 

\subsection{Graph creation}

To create graphs from a cell by gene counts feature matrix, we used a batch-balanced kNN graph~\cite{polanski_bbknn_2019}. BB-kNN constructs a kNN graph that identifies the top $k$ nearest neighbors in each "batch", effectively aligning batches to remove bias in cell source while preserving biological variability~\cite{bbknn_support}. We used annoy's method of approximate neighbor finding with a Euclidean distance metric in $50$-dimensional PCA space. Per "batch" we find $k=3$ top nearest neighbors. An example BB-kNN graph is schematized in main text, Figure 1A.  
\begin{figure}[h]
  \centering
  \includegraphics[width=\columnwidth]{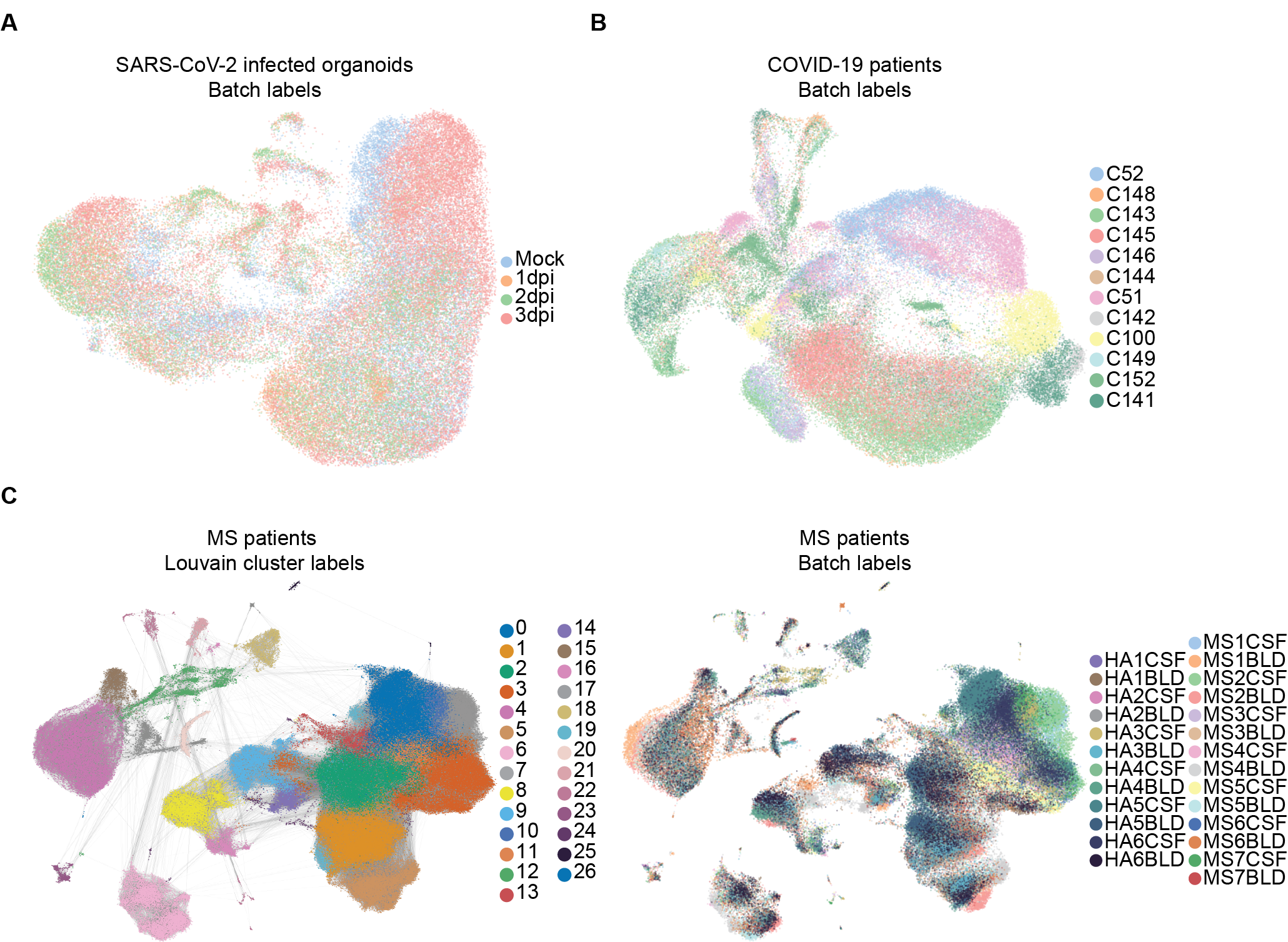}
  \caption[0.95\textwidth]{UMAP embeddings of individual cells colored by labels for auxiliary tasks. (\textbf{A}) Batch labels for SARS-Cov-2 infected organoids dataset. (\textbf{B}) Batch labels for COVID-19 patients, for patient IDs described in~\cite{liao_single-cell_2020}. (\textbf{C}) Graph used for the MS dataset with Louvain cluster labels to represent cell types (left) and batch labels per patient sample (right).}
  \label{fig:cell_type}
\end{figure}


\section{Hyperparameters and Training details}
\begin{table}[h]
\centering
\caption{Default hyperparameters used in the experiments}
\resizebox{\columnwidth}{!}
{
\begin{tabular}{ccc}
\hline
 & Graph Attention Network & Graph Convolution Network \\
\hline
Number of layers & $2$ & $2$ \\

Hidden\_size & $8$ & $256$ \\

Attention Heads & $8$ & N/A \\

Optimizer & Adagrad & Adagrad\\


weight\_decay & $.0005$ & $.0005$ \\

Batch size & $256$ & $256$\\

Dropout & $.5$ & $.4$ \\

Slope in LeakyRelu & $.2$ & $.2$ \\

Training Epochs & $1000$ & $1000$ \\

Early stopping &  $100$ & $100$\\
\hline
\end{tabular}

\label{tab:hyperparameters}
}
\end{table}

For auxiliary tasks and for training our models, we break our graph into $5000$ subgraphs using the ClusterData function in PyTorch Geometric library and then minibatched the subgraphs using the ClusterData function. These algorithms are originally introduced in~\cite{clusterloader}. We used a single block of Set Transformer with input dimension $18$, output dimension $8$ and $2$ heads. The rest of the hyperparamaters of GAT and GCN can be found in table~\ref{tab:hyperparameters}. 

For our auxiliary tasks and for baseline experiments we used an 8GB Nvidia RTX2080 GPU and for our main tasks we used an Intel E5-2660 v3 CPU with 121GB RAM.













\section{Auxiliary task}
In this section we describe our auxiliary tasks. Table~\ref{tab:aux_label} gives details about the number of labels for the auxiliary tasks. We first predict the cell types as given by the Louvain clustering~\cite{louvain}. In figure~\ref{fig:cell_type} we show Louvain community ID or cluster labels for the MS patients dataset, which can be annotated as cell types as in~\cite{ms_paper}. In the main text, we used~\cite{marker_db} to obtain cell type markers and annotate the Louvain cluster labels as "cell types" explicitly. 

Next we predict the batch ID of each node, i.e. which patient or from where the cell is obtained. Table~\ref{tab:aux_task} shows our results for these auxiliary tasks. 
In single-cell RNA-sequencing, variability between
batches can explain more of the transcriptomic variability than variability in the biological process of interest; these "batch effects" can complicate model inference~\cite{rnacluster}. Our novel use of BB-kNN graphs for the tasks described in this paper limits this "batch effect" bias. 


\begin{table}[h!]
\centering
\caption{Number of labels for auxiliary tasks}
\label{tab:aux_label}

\resizebox{\columnwidth}{!}{%
\begin{tabular}{l c c c c} 
\toprule
Task & \begin{tabular}[c]{@{}c@{}}SARS-CoV-2\\ infected organoids\end{tabular} & COVID-19 patients & Multiple sclerosis blood \& CSF &  \\ \midrule
Cell type  &      8         &        10           &     27           &  \\
Batch  &   4    &   12   &  26  \\
\bottomrule
\end{tabular}
}
\end{table}

\begin{table}[h!]
\centering
\caption{Results on auxiliary tasks}
\label{tab:aux_task}

\resizebox{\columnwidth}{!}{%
\begin{tabular}{l c c c c} 
\toprule
Prediction & \begin{tabular}[c]{@{}c@{}}SARS-CoV-2\\ infected organoids\end{tabular} & COVID-19 patients & Multiple sclerosis blood \& CSF &  \\ \midrule
Cell type  &      93.84         &        82.03           &     75.90           &  \\
Batch  &   76.16    &   64.08   &  36.21  \\
\bottomrule
\end{tabular}
}
\end{table}

\section{Code and Data Availability}
The processed data for SARS-CoV-2 infected organoids samples and the COVID-19 patient samples can be found at~\href{https://drive.google.com/drive/folders/10YLEJfM-GL1XTZj1Sjs4z9b0SmELyMUp?usp=sharing}{this link}. We do not have the permission to share the MS patient data, as it belongs to a third party. All code used to reproduce these results are available in the associated supplementary material and will be published on GitHub post-review.  



\end{document}